# Impact of magnetic domains on magnetic flux concentrators

Federico Maspero[1], Simone Cuccurullo[2], Dhavalkumar Mungpara[3], Alexander Schwarz[3] and Riccardo Bertacco[1,2]

[1]CNR Istituto di Fotonica e Nanotecnologie, Milano, Italy

[2]Politecnico di Milano, Milano, Italy and

[3]Institute of Nanostructure and Solid State Physics, University of Hamburg, Germany.

## Abstract

The impact of magnetic domains in magnetic flux concentrators is studied using the simulation software MuMax3. First, the simulation parameters are validated using experimental results from magnetic force microscopy; second the simulation output is benchmarked with the one obtained using Comsol Multiphysics. Finally, the impact of magnetic domain is assessed, showing how micromagnetic effects can become relevant, if not dominant, when scaling the gap between the MFC and the sensor.

## 1 Introduction

Integrated magnetic flux concentrators (MFCs) are made of thin films of ferromagnetic materials with high permeability and low coercive field, e.g. permalloy ($Ni_{80}Fe_{20}$) or Co based alloys like CoZrNb/CoZrTa. They are used to amplify [1, 2] and/or modulate [3] an external magnetic field nearby a magnetic field sensor, e.g. a magneto-resistive sensor, which is placed in the gap between two MFCs.

Conventionally, the design of MFCs is optimized by solving the magnetostatic Maxwell's equations through finite element methods [4, 5]. This approach allows to simulate relatively large geometries (MFCs have generally planar dimensions of hundreds of µm) but neglects micromagnetic effects, i.e. formation of magnetic domains and magnetic domain walls, which can be important at the µm scale.

The study of the effect of micromagnetic domains on MFC field amplification was partially assessed in the work of Trindade et al. [2, 6], yet no modelling or tool to study this effect was given. In this paper, therefore, micromagnetic simulations and experimental measurements are performed to understand the effect of magnetic domains on the amplification behaviour of different MFC shapes.

The paper is divided as follows: first, a description of the employed simulation/experimental methods, specifying the adopted parameters. Second, the comparative investigation, both by magnetic force microscopy (MFM) and by simulations (using MuMax3) of the micromagnetic configuration vs. applied field for rectangular MFC, to tune the parameters to be used for micromagnetic simulations [7]. Finally, the comparison between magnetostatic and micromagnetic simulations for various shapes of MFC, to assess the effect of magnetic domains on the field amplification curve of MFC.

## 2 Simulation Methods

### 2.1 MuMax3

Micromagnetic simulations of the MFCs were performed with MuMax3 [7], which is a GPU-accelerated software that offers a significant speedup compared to other CPU-based computer programs.

MuMax3 employs the finite difference method to simulate the time and space dependent magnetization evolution at the micrometric scale by solving the Landau-Lifshitz-Gilbert equation [8, 9]:

$$\frac{\partial \boldsymbol{m}}{\partial t} = \gamma \boldsymbol{H}_{eff} \times \boldsymbol{m} + \alpha \boldsymbol{m} \times \frac{\partial \boldsymbol{m}}{\partial t}$$

where $\boldsymbol{m}$ is the reduced magnetization, $\gamma$ is the gyromagnetic ratio, $\alpha$ is the damping constant and $\boldsymbol{H}_{eff}$ is the effective field comprising the exchange, the Zeeman and the demagnetizing contributions.

Magnetic parameters of permalloy are employed, i.e. saturation magnetization $M_s = 8 \times 10^5$ A/m, exchange stiffness $A = 1.0 \times 10^{-12}$ J/m and null anisotropy constant. The typical damping constant for permalloy is $\alpha = 0.01$. However, since only stationary conditions are considered, $\alpha$ was set to 0.1 to speed up the simulations.

The size of the discretization cell is bounded by the exchange length $\lambda_{ex} = (2A/\mu_0 M_s^2)^{1/2} \approx 5$ nm. Nevertheless, because of the relatively large geometries simulated in this work, cells sizes ranging from 8 to 25 nm have been used. This choice results in a loss of resolution for magnetic domain walls, which is not critical for the scope of the paper as confirmed by experimental results.

### 2.2 COMSOL Multiphysics

Magnetostatic simulations of the MFCs were performed with COMSOL Multiphysics by solving the Maxwell's equations in stationary conditions. The simulations comprised an air volume at whose centre are placed two MFCs separated by a gap, immersed in a uniform magnetic field. Both air and MFCs were modelled as linear and isotropic magnetic materials with relative permeability equal to 1 (air) and 1000 (MFCs). The latter was chosen in accordance with measurements on permalloy thin films used in this work and those found in literature [10].

# 3 Experimental method

## 3.1 Fabrication

Magnetic field concentrators were patterned through optical lithography on a 1 x 1 cm$^2$ silicon substrate. The resist adopted was the AZ 5214 E image reversal photoresist (Microchemicals).
Permalloy was deposited by electron beam evaporation (Evatec BAK 640). The base pressure was 10$^{-7}$ mbar and the deposition rate was set to 0.2 nm/s.

## 3.2 MFM characterization

We performed magnetic force microscopy (MFM) measurements [11] with a commercial instrument [12] in ambient conditions using amplitude modulation [13] and the lift mode technique. In the MFM lift mode, the tip is moved at a constant height above the surface without ever touching the surface by following the topography path recorded during the first scan. The phase $\varphi(x,y)$ is recorded, which reflects the long-range magnetostatic tip-sample interaction and is called MFM image.

We used commercially available CoCr coated tips with a spring constant $c_z$ = 2.8 N/m and a resonance frequency $f_r$ = 69.3 kHz [14]. Due to the shape anisotropy energy of the thin magnetic film on the tip pyramid, the tip magnetization is oriented nearly perpendicular to the sample surface [15]. This results in a predominantly out-of-plane sensitivity [16]. For samples like permalloy thin films with in-plane magnetized domains, out-of-plane components are only present at domain walls. The situation is sketched in Fig. 1a-b for Bloch and Neel type 180° domain walls, in which the magnetization rotates out-of-plane and in-plane, respectively. The characteristic contrast change across a domain wall allows identifying the type of domain wall present in the sample.

To apply in-plane magnetic fields of up to 20 mT, we used a calibrated electromagnet (two Cu coils with yokes in a Helmholtz configuration). Even at the maximum field, the tip-magnetization remained stable and only the domain structure of the permalloy sample changed.

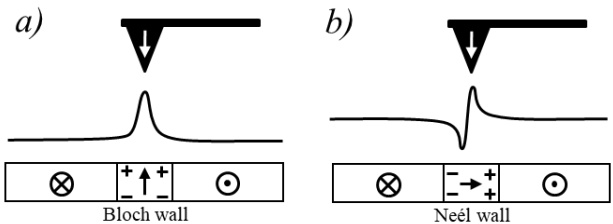

*Figure 1 – a)-b) MFM contrast across (a) Bloch and (b) Neél-type domain wall. Shape anisotropy energy forces the tip magnetization to be nearly normal to the sample surface. Therefore, in MFM images Neél walls can be easily identified by a dark-bright contrast pattern between neighboring magnetic domains.*

# 4 Results

## 4.1 Investigation of rectangular MFC by MFM and MuMAx3 simulations

Fig. 2 and Fig. 3 show the comparison between the experimental results obtained by MFM and micromagnetic simulations by MuMax3 for rectangular MFC, with 20 x 26 mm area and 150 nm thickness. It should be pointed out that two different physical quantities are represented in Fig.2 and Fig.3. The result of the simulation is shown by plotting the x-component of the computed magnetization. This gives an idea of the domain configuration and highlights the component which is relevant for the calculation of field amplification produced by the MFC. On the other hand, the MFM measurement shows the interaction force between the probe tip and the stray field of the sample in the out-of-plane direction. This gives information on the presence of domain walls and the type of wall, but it does not directly shows the magnetization of the domain. Nevertheless, the extrapolated domain map and its evolution under applied field can be used for comparison of the two methods.

Experimentally we found two characteristic micromagnetic flux-closure configurations at zero applied field: (i) a Landau pattern with two vortexes and a 180° central domain wall (Fig. 2d), (ii) a diamond pattern with a central hexagonal domain and four vortices (Fig 3d). The magnetization direction within the domain are indicated by arrows on the zero-field configurations. All 90° and most 180° domain walls are of Nèel-type, clearly identifiable by the dark-bright contrast between neighbouring domains. However, sometimes we also observe more complex 180° cross-tie walls. MFM results show that the initial domains configurations can change from one MFC to another likely due to presence of defects in the film. In any case, applying an external field always increases the size of those domains, which are parallel to the field direction. In both cases, an external magnetic flux density of 10 mT does not fully saturated the elements and resultes in an S-state.

To simulate the behaviour of MFCs with a Landau pattern the zero-field magnetization of the MFC was obtained upon relaxation of a single central vortex magnetic configuration, which is typically found in rectangular dots of Permalloy of this size [17].

After reaching the equilibrium state an external field in the x direction ranging from 1 to 10 mT was applied in simulations, together with a cross-axis field of 1 % both in y and z direction to avoid instabilities of the solution and account for misalignment in the experimental setup. The equilibrium magnetization was recorded at each step and compared with the experimental results, as shown in Fig. 2 and Fig. 3. In both experiment and simulations, the same motion of domain walls is observed, corresponding to the expansion of the left domain aligned to the external field.

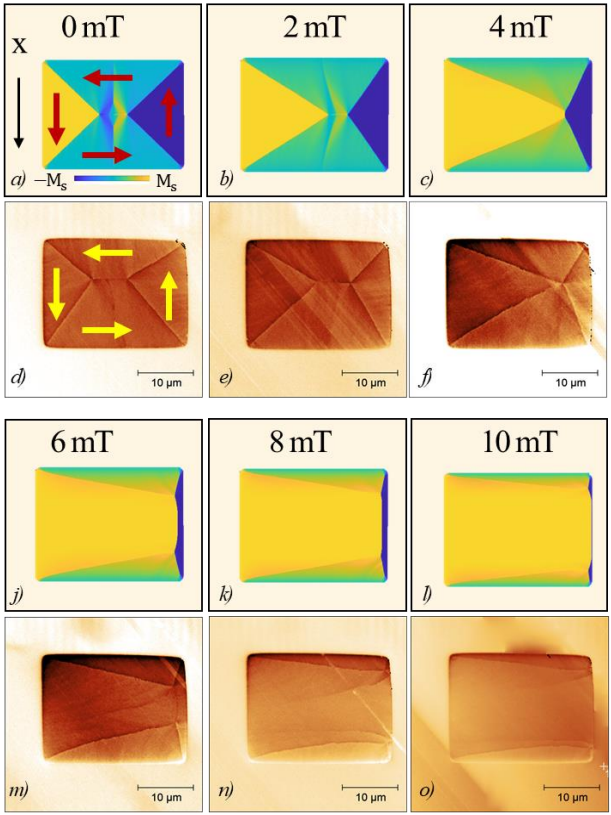

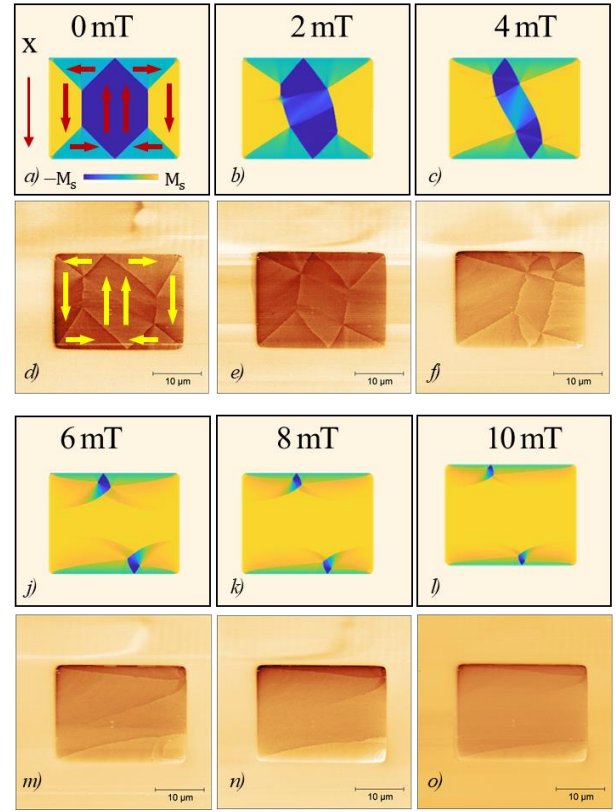

*Figure 2 – MFM measurement (bottom images) and x-axis component of the simulated magnetizaton (top images). Mumax simulation were performed with a 1024 x 1024 x 16 cells grid, the size the rectangular shape is 20 x 26 x 0.15 μm. This leads to a miximum cell size of 25nm. The magnetization was acquired in the central plane of the simulated film, however the same domains configuration was observed in all horizontal planes.*

Good agreement among the simulation and the experimental results can be seen also in the case of the diamond configuration. This configuration was less common in the array of fabricated permalloy MFC, but still present. Such configuration was reproduced in MuMax3 by placing two magnetization vortexes with anticlockwise and clockwise rotation in the left and right region of the MFC respectively. This time, domain walls motion takes place together with coherent rotation and domain reconfiguration.

Overall, this comparative study allowed to tune the simulation parameters (see methods) and find a very good agreement between simulations and MFM experiments. The very same parameters have been used also for other geometries, to assess the impact of the MFC micromagnetic structure on their functional properties.

*Figure 3 – Sequence of MFM images and MuMax simulation results starting with two vortexes configuration. With no applied field the two vortexes relax into an exagonal central domain. The applied field has the same direction of the external domains magnetization. This causes the centreal exagon to shrink until a large central domain is formed due to the opposite direction of field and magnetization. The two domains colored in blue which are evident in the simulation are not so clearly visible in the experimental measurement where they progressivly move to the boarder of the MFC and disappear.*

### 4.2 MuMax3 as a tool to model MFC

In the second part of this work, MuMax3 and COMSOL were used to model the magnetization of three different geometries of MFC under varying magnetic field and evaluate the demagnetizing field produced by the three shapes. The three chosen geometries are typical MFC geometries [4]: funnel-like, T-shape and rectangular.

The size of the MFC (5x8 micron) was chosen to be smaller than those used for the previous simulation (Fig. 2 and Fig. 3) to save computational time and simulate pairs of MFCs instead of a single MFC.

Fig. 4 shows an example of the magnetization map obtained with the two softwares.

The x component of the magnetization obtained with COMSOL Multiphysics follows the results observed in literature [5]: the magnetization reaches the maximum value in the center of the MFC and decreases towards the edges. A different result is found in MuMax3, where the magnetization map shows magnetic domains. Again, a vortex-like configuration is seen in both concentrators, with the upper and lower parts showing opposite magnetization. However, in the left one there is just a 180° DW, while in the right-one a diamond-like configuration appears. The difference between the left and right concentrators depends on the sweep order of the field. Indeed, the same simulation with opposite sweep order (+80 mT to -80 mT instead of -80 mT to +80 mT) leads to an almost specular result (Figure 4b-c). This is due to the asymmetric shape of the MFC along x, which together with the applied field direction, has an impact on the magnetic domains formation.

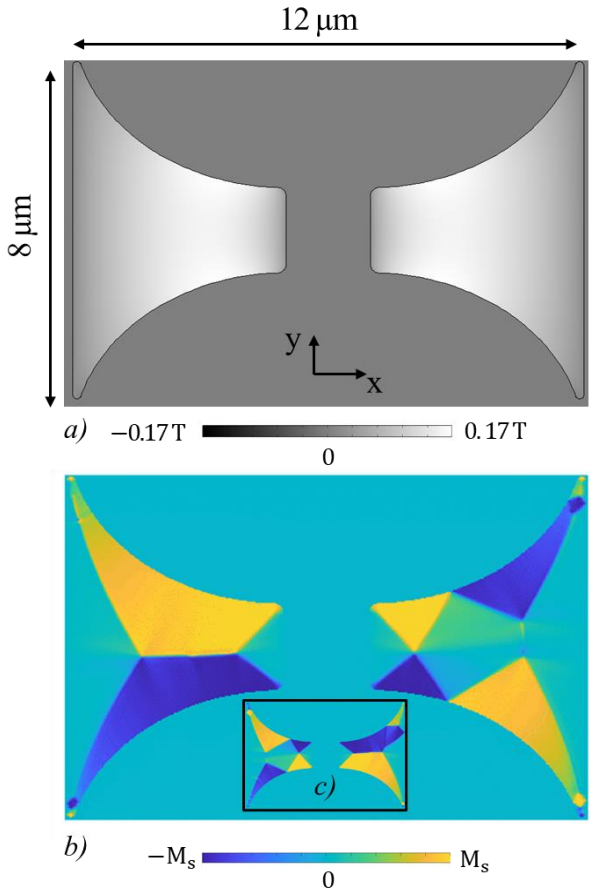

*Figure 4 – Magnetization ($M_x$) of the funnel-like geometry computed with COMSOL (a) and MuMax3 (b-c) when applying 2 mT in the x direction. This time MuMax simulation was started from a saturated magnetization and then the magnetic field was swept from -80 mT to +80 mT and viceversa. Figure b) and c) show the magnetization map obtained under 2 mT for different sweep order. The magnetization asymmetry of the MFC is mirrored when the applied field sweeping order is reversed. The grid was set to 1024 x 1024 x 16. The largest cell size was around 12 nm. Scale is reported for size evaluation. The thickness was set to 150 nm as in the experimental case. Comsol magnetization is gradually increasing towards the center of the geometry and it is linearly proportional to the external field. MuMax shows a more realistic domains magnetization.*

Note, however, that here what is crucial is the magnetization close to the central poles, in a region where left and right MFCs have a similar domain.

The second crucial aspect studied in the simulations is the field produced in the region between the two MFC. Fig. 5 compares the demagnetizing field vs. applied external field at the center of the gap. The results obtained with the two simulation tools agree only in the central region. The effects of DW motion under external field and magnetic saturation appears only in micromagnetic simulations, clearly showing deviations from the linear behaviour at a few mT.

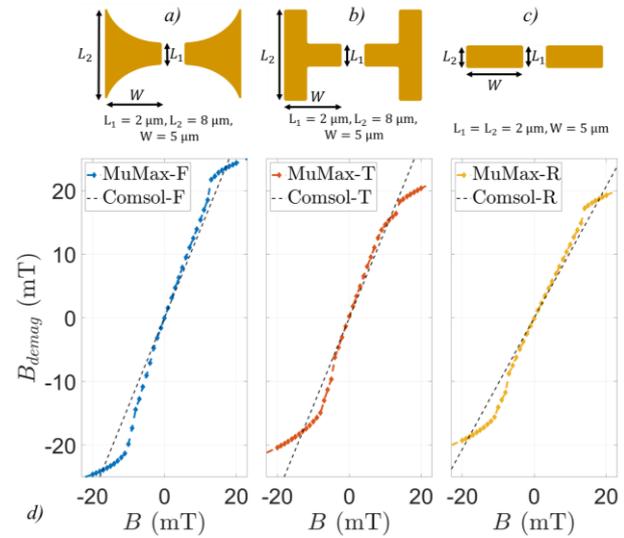

*Figure 5 – d) Demagnetizing field versus applied field at the center of the constriction computed using COMSOL and MuMax3 for funnel-like (a), T-shaped (b) and rectangular (c) MFCs. The field was averaged over a rectangular area in the center of the constriction. x=100nm and y=2um. Only half hysteresis cycle (sweep-up) is shown to better compare the slope of the curves.*

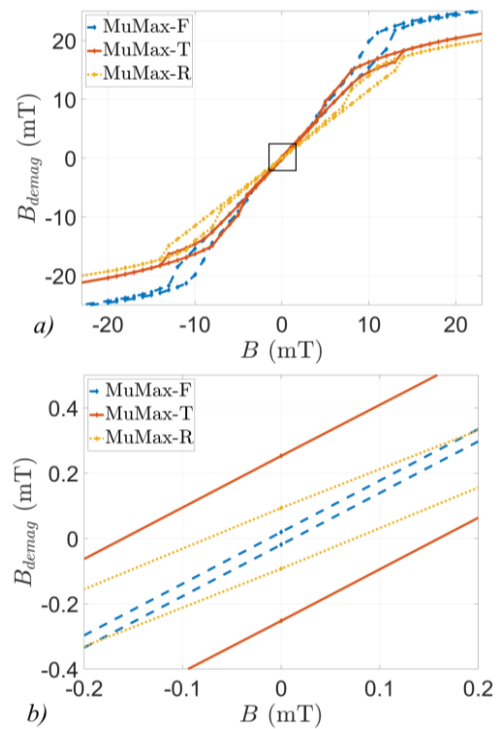

*Figure 6 – a) Comparison between the hysteresis cycles of the three MFC shapes. The simulation was computed only by sweeping the field from -80 to 80 mT and assuming a symmetric response of opposite sign for the down-sweep. b) zoomed image around zero applied field to evaluate the coercivity and the slope of the three curves.*

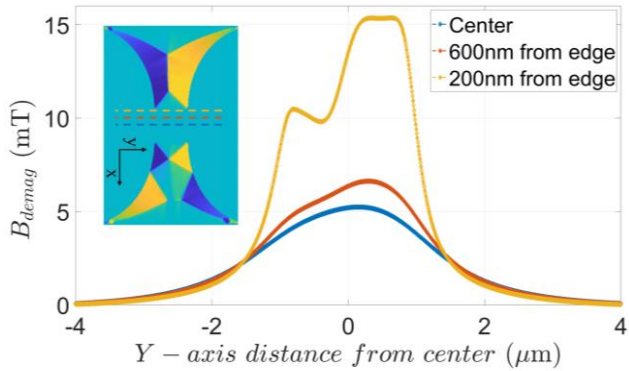

*Figure 7 – Demagnetizing field produced by the MFC along y-axis cutline taken at different distances from the center (represented in figure insert). The graph is obtained with 2 mT applied. It is visible how the demagnetizing field is related to the domains configuration. First, the maximum is not found at the center of the cutline. Second, close to the MFC the field distribution shows lobes which depend on the magnetic domain distribution.*

In Fig. 6 the simulated hysteresis cycles of the three MFC shapes are reported.

The slope of the curves for small fields, i.e. the MFC amplification factor, increases proportionally to the ratio $L_2/L_1$ (see Fig. 5). Therefore, the funnel-like and T-shape geometries have the largest amplification factor.

Furthermore, the funnel-like and rectangular geometries show smaller coercivities. This is likely due to the absence of pinning sites for magnetic domains, which are present in the T-shape MFC for geometric reasons.

The funnel-like configuration thus presents the best characteristics in terms of coercivity, amplification factor and linear range.

The final aspect which was investigated was the effect of the domains on the uniformity of the field produced by the magnetic flux concentrator. In Fig. 7, the value of the demagnetizing field along the y-axis is plotted at different distances from the center of the constriction.

As it can be seen, when moving close to the MFC edges, it is important to consider that the field produced by the MFC is not uniform and related to the domain distribution. For instance, the sensitivity of a sensor with a square active area of 1x1um could be affected by misalignment error in the fabrication process. The same sensor at different y-axis coordinate would respond differently to an external field.

## 5 Conclusions

Magnetic flux concentrators are magnetic objects used to amplify and shape the magnetic field. They are currently used in several magnetic sensors.

In this work, different MFCs geometries were studied considering micromagnetic effects, in particular the fragmentation into magnetic domains.

First, the parameters to be used in the micromagnetic simulation tool (MuMax3) were tuned by comparison with experimental result from MFM measurement of permalloy MFC under applied field.

Second, the field amplification of three different MFC geometries was studied using MuMax3 and COMSOL Multiphysics. Our results show good agreement between the simulation platforms only in the linear regime, i.e. within a few mT. With respect to COMSOL, MuMax3 enables the possibility of studying micromagnetic effects giving more realistic modelling of the MFC behavior, including hysteresis, saturation and impact of the micromagnetic configuration on the spatial profile of the field within small gap MFCs.

On the other hand, COMSOL offers a much faster simulation tool and is more suited for simulation of large size MFC (hundreds of microns).

## Acknowledgments

This work was carried out in the frame of the European FET-Open Project OXiNEMS. This project has received funding from the European Union's Horizon 2020 research and innovation programme under grant agreement No 828784.

## Open data

The data that support the findings of this study and supplementary material are openly available in a Zenodo repository with DOI:10.5281/zenodo.4446957.